\documentclass[a4paper,onecolumn]{IEEEtran}
\usepackage{algpseudocode}
\usepackage{algorithm}
\usepackage[tbtags]{amsmath}
\usepackage{amsfonts}
\usepackage{amssymb}
\usepackage{amsthm}
\usepackage{cite}
\usepackage{calc}
\usepackage{color}
\usepackage{epsfig}
\usepackage{setspace}
\usepackage{epstopdf}

\newtheorem{defn}{Definition}
\newtheorem{thm}{{\cal T}heorem}[section]
\newtheorem{cor}[thm]{Corollary}
\newtheorem{prop}{Proposition}
\newtheorem{lem}[thm]{Lemma}
\newtheorem{conj}[thm]{Conjecture}
\newtheorem{constr}[thm]{Construction}
\newtheorem{note}{Remark}
\newcommand{\bit}{\begin{itemize}}
\newcommand{\eit}{\end{itemize}}
\newcommand{\bcor}{\begin{cor}}
\newcommand{\ecor}{\end{cor}}
\newcommand{\beq}{\begin{equation}}
\newcommand{\eeq}{\end{equation}}
\newcommand{\beqn}{\begin{equation}}
\newcommand{\eeqn}{\end{equation}}
\newcommand{\bea}{\begin{eqnarray}}
\newcommand{\eea}{\end{eqnarray}}
\newcommand{\bean}{\begin{eqnarray*}}
\newcommand{\eean}{\end{eqnarray*}}
\newcommand{\ben}{\begin{enumerate}}
\newcommand{\een}{\end{enumerate}}
\newcommand{\bdefn}{\begin{defn}}
\newcommand{\edefn}{\end{defn}}
\newcommand{\bnote}{\begin{note}}
\newcommand{\enote}{\end{note}}
\newcommand{\bprop}{\begin{prop}}
\newcommand{\eprop}{\end{prop}}
\newcommand{\blem}{\begin{lem}}
\newcommand{\elem}{\end{lem}}
\newcommand{\bthm}{\begin{thm}}
\newcommand{\ethm}{\end{thm}}
\newcommand{\bconj}{\begin{conj}}
\newcommand{\econj}{\end{conj}}
\newcommand{\bconstr}{\begin{constr}}
\newcommand{\econstr}{\end{constr}}
\newcommand{\bpf}{\begin{proof}}
\newcommand{\epf}{\end{proof}}
\newcommand{\bprf}{{\em Proof: }}
\newcommand{\eprf}{\hfill $\Box$}

\begin{document}

\title{Determining the Generalized Hamming Weight Hierarchy of the Binary Projective Reed-Muller Code}
\author{
	\IEEEauthorblockN{ Vinayak Ramkumar, Myna Vajha and P. Vijay Kumar}\\
	\IEEEauthorblockA{Department of Electrical Communication Engineering, Indian Institute of Science, Bangalore.\\
		Email: \{vinayak, myna, vijay\}@iisc.ac.in} 
	\thanks{Myna would like to thank the support of Visvesvaraya PhD Scheme for Electronics \& IT awarded by DEITY, Govt. of India. P. Vijay Kumar is also a Visiting Professor at the University of Southern California.  His research is supported in part by the National Science Foundation under Grant No. 1421848 and in part by the joint UGC-ISF research program.}
}
\maketitle

%\ \\
%\bit
%\item Abstract
%\item Introduction
%\bit
%\item What is ghw and viktor wei work on RM code
%\item projective reed muller codes 
%\item mention works on ghw, weight enumerators (Carvello)
%\item Description of current work that includes shortening algorithm, upper bounds and claiming the ghw by proving a lower bound based on Viktor Wei's method.
%\eit
%\item Section 1: PRM (Sub code of RM code)
%\item Section 2: Shortening Algorithm (Upper Bounds on GHW) (Example)
%\item Section 3: Lower Bounds on GHW of PRM
%\item Section 4: GHW of PRM codes. (Graph for PRM and RM GHW)
%\item Conclusions
%\eit

\begin{abstract}
%In this paper, we determine the complete GHW hierarchy of the binary, projective Reed-Muller (PRM) code. 
Projective Reed-Muller codes correspond to subcodes of the Reed-Muller code in which the polynomials being evaluated to yield codewords, are restricted to be homogeneous. The Generalized Hamming Weights (GHW) of a code ${\cal C}$, identify for each dimension $\nu$, the smallest size of the support of a subcode of ${\cal C}$ of dimension $\nu$.   The GHW of a code are of interest in assessing the vulnerability of a code in a wiretap channel setting.  It is also of use in bounding the state complexity of the trellis representation of the code. 

In prior work~\cite{VajRamKum} by the same authors, a code-shortening algorithm was employed to derive upper bounds on the GHW of binary projective, Reed-Muller (PRM) codes.   In the present paper, we derive a matching lower bound by adapting the proof techniques used originally for Reed-Muller (RM) codes by Wei in \cite{Wei}. This results in a characterization of the GHW hierarchy of binary PRM codes.
\end{abstract}

\section{Introduction}
The notion of Generalized Hamming Weights (GHW), introduced by Wei in \cite{Wei}, is a  generalization of minimum Hamming weight of a linear code. In \cite{Wei} , the basic properties of GHW  are studied and the weight hierarchy for Hamming code, Reed-Solomon codes, binary Reed-Muller code  etc are determined.

Study of this notion was motivated by applications in cryptography. For instance, when a linear code is used over wire-tap channel of type II (see \cite{OzarowWyner}), the amount of information revealed can be completely characterized using GHW hierarchy of the linear code. In a similar way, GHW can be used to analyze the performance of a linear code when used as a $t$-resilient function \cite{ChorGoldBit}. Later, study of GHW hierarchies found applications in determining the optimum bit order in trellis based decoding. Specifically, the GHW of RM codes found in \cite{Wei} were used in \cite{kasami1993optimum} to prove that standard binary bit order is optimal for RM codes. 

%
%construction of resilient functions {\color{red} expand wire tap, add trellis complexity analysis}. 

The $\nu$-th GHW of a code $\mathcal{C}$ is given by,
\bean
d_{\nu}(\mathcal{C}) = \min \big\{ \ | S(D)| : D \ \text{is a subcode of} \ C \\ \text{with dimension} \ \nu \big\},
\eean
where $\text S(\mathcal{D})$ denotes the union of support of all the vectors in $\mathcal{D}$.

%The knowledge of GHW of a code is useful in analyzing problems involving the shortened versions of that code. 
 A geometric approach to determine GHW hierarchy for various classes of codes is described in \cite{TsfVla}. The projective Reed-Muller (PRM) codes, introduced by Gilles Lachaud in \cite{Lachaud86}, are a variant of the Reed-Muller (RM) codes. These codes are based on evaluations of homogeneous polynomials of degree $r$ in projective space $\mathbb{P}^{m-1}(\mathbb{F}_q)$.  The dimension and minimum distance of  PRM codes were determined by Serre \cite{serre} and S$\o$rensen  \cite{Sorensen}. In \cite{boguslavsky}, Boguslavsky determined the second GHW of projective Reed-Muller codes for $r < q-1$ regime. The connections between Tsfasman-Boguslavsky conjecture and GHW of PRM codes were studied in \cite{DattaGhorpade}. To the best of our knowledge, none of the previous works on GHW of PRM codes have considered the binary ($q=2$) version. However, next-to-minimal weight of binary PRM codes is determined in a recent work \cite{CarvalhoNeumann}, wherein next-to-minimal weight means minimal codeword weight that is greater than minimum Hamming weight. Note that the next-to-minimal weight is not the same as second generalized Hamming weight.

In this paper we present the GHW hierarchy for  binary PRM codes. In a recent work \cite{VajRamKum}, by authors of the current paper, it was shown that binary PRM codes and their shortened versions have Private Information Retrieval (PIR)  code property. The shortening technique proposed in that work resulted in an upper bound on the GHW of binary PRM codes. The work presented in this paper started as an attempt to prove the optimality of that shortening procedure. Here we derive a lower bound for the GHW of PRM codes and show that it matches with the upper bound provided in \cite{VajRamKum}. The proofs presented in this paper adapt the ideas from derivation of GHW for RM codes in \cite{Wei}. 

\paragraph{Organization of paper}In Section~\ref{sec:PRM} we describe the parameters and properties of binary, projective Reed-Muller code. Section~\ref{sec:short} discusses about the shortening procedure proposed in  \cite{VajRamKum} that gives an upper bound on GHW of  binary PRM codes. A lower bound is derived in Section~\ref{sec:lower} using techniques from \cite{Wei}. In Section~\ref{sec:ghw} we show that these bounds match and thereby determine the GHW hierarchy of the binary projective Reed-Muller code.
\paragraph{Notation} We use $d_{\nu}(r,m)$ to denote the $\nu$-th GHW for PRM$(r,m-1)$ code.  The notation $[a,b]$  denotes $\{a,a+1, \cdots ,b-1,b\}$ and $[a]=[1,a]$. The support of any code $\mathcal{C}$ is denoted by $S(\mathcal{C})$.  

\section{Binary Projective Reed-Muller Codes}
\label{sec:PRM}
%The Projective Reed-Muller (PRM) codes were introduced by Gilles Lachaud  \cite{Lachaud86} and are based on homogeneous multivariate polynomials. These codes are  a variant of the classical Reed-Muller Codes.
Every codeword in the $\text{PRM}(r, m-1)$ code over the field $\mathbb{F}_q$ is a vector of evaluations of a homogeneous polynomial of degree $r$ at a fixed representative of each of the points in the projective space $\mathbb{P}^{m-1}(\mathbb{F}_q)$.

For $q=2$, each point in the projective space $\mathbb{P}^{m-1}(\mathbb{F}_2)$ has a unique representative with $m$ components, i.e,
$\mathbb{P}^{m-1}(\mathbb{F}_2) \simeq \mathbb{F}_2^m \setminus \{\underline{0}\}.$
A codeword in binary PRM$(r,m-1)$ code is the vector of  evaluations at all non-zero points in $\mathbb{F}_2^{m}$ of a binary homogeneous polynomial of degree $r$ in $m$ variables.
\bea
\label{eq:PRM}
f(\underline{x})=f(x_m, \cdots, x_1) = \sum\limits_{|R| = r, R \subseteq [m]} a_{R} \prod\limits_{i \in R} x_i
\eea
where $a_R \in \mathbb{F}_2$.
The coefficients of monomials $\{a_R,|R|=r\}$ represents the message symbols. It can be easily seen that binary PRM code is a systematic code.
Here we will be discussing only about binary PRM codes and hence from now on-wards PRM would mean the binary version.  

Any binary homogeneous polynomial of degree $r$ evaluates to $0$ at vectors (in $\mathbb{F}_2^{m}$) with Hamming weight less than $r$. Hence, there are some coordinates which are always zero in all the codewords and can be deleted from the binary $\text{PRM}(r, m-1)$ code. 
The \emph{non-degenerate} $\text{PRM}(r,m-1)$ code thus obtained has parameters:
\bea
\text{Dimension \ }  &=& {m \choose r}, \nonumber \\
\text{Block length \ } &=&2^m-\sum\limits_{i = 0}^{r-1} {m \choose i}= \sum\limits_{i = r}^m {m \choose i}.
\eea
\paragraph*{Support-set viewpoint} Note that each codeword of this code is of the form $(f(\underline{x})$,  $\underline{x} \in \mathbb{F}_2^m$ with $w_H(\underline{x}) \ge r)$. Any vector $\underline{x} \in \mathbb{F}_2^m$ can be represented uniquely by its support. This implies that each code symbol can  be indexed by a subset of $[m]$ with size $\ge r$.  For an example code with $m=4$,  code symbol $f(1011)$ can be represented as $f(\{1,3,4\})$ where $\{1,3,4\}$ is the support of vector $(1011)$. 

\begin{note}
	Each message symbol as well as its corresponding monomial can be indexed by a $r$-element subset of $[m]$. 
\end{note}

\section{Shortening Algorithm : Upper Bound}
\label{sec:short}
In this section, we briefly describe the shortening technique proposed in \cite{VajRamKum}, that resulted in upper bound on GHW of binary PRM codes. 

For the $PRM(r,m-1)$ code, any code-symbol $f(S), S \subseteq[m]$, is given by:
\bean
f(S) = \sum\limits_{R_i \subseteq S} a_{R_i},
\eean
where $R_i, \forall i \in \left[ {m \choose r} \right] $ are the $r$-element subsets of $[m]$ and $a_{R_i}$ are the message symbols. 

For an example code with $r=2$ and $m=4$, $f(\{1,3,4\}) = f(\{1,3\}) + f(\{1,4\}) + f(\{3,4\})$.\\

Consider that we set the message symbols $a_{R_i}=0$, $\forall R_i \subseteq S$. This is equivalent to setting $f(R_i) = 0$, $\forall R_i \subseteq S$ because the code is systematic. Now we have, 
\bea 
f(S) = \sum\limits_{R_i \subseteq S} f(R_i)= 0.
\eea
This means that the coordinates corresponding to $\{R_i \ | \ \forall R_i \subset S, i \in {m \choose r} \} \cup \{S\}$ can be ignored. Hence, on shortening the PRM$(r,m-1)$ code by setting all message symbols corresponding to some $\gamma$ $r$-element subsets to zero, %of a set $S \subseteq [m]$ to zero, 
we can ignore the code coordinates corresponding to the message symbols and possibly some other code coordinates. % that have all the subsets included in the picked message symbols.
Therefore, this shortening procedure will result in block length reduction of $\Gamma(r,m,\gamma) \ge \gamma$. The resultant code obtained will have parameters: 
\bea
\text{Dimension } k &=& {m \choose r} - \gamma,  \nonumber \\
\text{Block length } n  &=& \sum\limits_{i = r}^m {m \choose i} - \Gamma(r,m,\gamma).
\eea
The aim of a good shortening algorithm for PRM$(r,m-1)$ code should be to pick  these message symbols ( $r$-element subsets of $[m]$ ) so that block length reduction is more. With this background, we state without proof the following lemmas  from \cite{VajRamKum}. For a given $\gamma$, $r$ and $m$, first a unique vector $\underline{\rho}$ is computed and then $\Gamma(r,m,\gamma)$ is computed using that. Note that $\ell=m-r$ here.  
\begin{lem}[ Unique $\underline{\rho}$ representation\cite{VajRamKum}~]
	\label{lem:unique}
	Any $\gamma  < {m \choose \ell}$ can be uniquely represented using a vector $\underline{\rho}=(\rho_{\ell-1}, \cdots \rho_0)$ with $\rho_i \ge 0, \forall i \in [0, \ell-1]$ and $\sum\limits_{i=0}^{\ell-1} \rho_i \le r$ as, 
	\bean 
	\gamma &=& \sum\limits_{t = 0}^{\ell - 1} h\left(\rho_t, r_t, t\right), \\ \text{ where } \ h\left(p, r, t\right) &=& \begin{cases}
		\sum\limits_{i = 0}^{p-1} { r+t-i \choose r-i} & p > 0\\
		0 & p = 0 
	\end{cases} \ \\ \text{ and } \ \  r_t &=& r - \small { \sum\limits_{q > t}^{\ell-1} \rho_{q}}.
	\eean
\end{lem}

\begin{lem}[ Block length reduction \cite{VajRamKum}~]
	\label{thm:shorten3} Let $\underline{\rho}=(\rho_{\ell-1}, \cdots \rho_0)$  be the unique representation of a given $\gamma \in \big[0, {m \choose \ell}\big)$. Let $r_t$, $\forall t \in [0, \ell-1]$, be as defined in the previous lemma.
	By setting $\gamma$ message symbols of \text{PRM}(r, m-1) code to zero, block length reduction of
	\bean 
	\Gamma(r,m,\gamma)  &=& \sum\limits_{t=0}^{\ell - 1} g\left(r_t, t\right), \\
	\text{ where } \  g\left(r, t\right) &=& \begin{cases}
		\sum\limits_{j = 0 }^{t} \sum\limits_{i=0}^{\rho_t - 1} {r+t-i \choose r+j-i} & \rho_t > 0\\
		0 & \rho_t = 0,
	\end{cases}
	\eean
	is possible. 
\end{lem}

The Table~\ref{table:PRM_2_4} shows the shortening procedure which results in Lemma~\ref{thm:shorten3} for  the case $m=5$, $r=2$. To reduce dimension by $\gamma$ one has to pick first $\gamma$ 2-element sets in the column $\mathbb{S}$ and set corresponding message symbols to zero. For example, if $\gamma=4$, the message symbols given by  $\{1,2\}$, $\{1,3\}$, $\{2,3\}$ and $\{1,4\}$ are set to zero.

\begin{table}[h!]
	\begin{center}	
		\bean
		\begin{array}{|c|c|c|c|c|}
			\hline k & \gamma  &  \mathbb{S} & \Gamma(2,5,\gamma) & n \\ \hline
			10 & 0 & \phi & 0  & 26\\ \hline
			9 & 1 &  \{1,2\} & 1 & 25\\ \hline
			8 & 2 &  \{1,3\} &  2 & 24 \\ \hline
			7 & 3 &  \{2,3\} & 4 &  22 \\ \hline
			6 & 4 &  \{1,4\} & 5 &  21\\ \hline
			5 & 5 &  \{2,4\} & 7 &  19\\ \hline
			4 & 6 & \{3,4\} & 11 &  15\\ \hline
			3 & 7 &  \{1,5\} & 12 &  14 \\ \hline
			2 & 8 & \{2,5\} & 14 &  12\\ \hline
			1 & 9 & \{3,5\} & 18 &  8 \\ \hline
		\end{array}
		\eean
		\caption{ Shortening procedure for PRM code with $r=2, m=5$.}% $\text{PRM}(2,4)$}
		\label{table:PRM_2_4}
	\end{center}
\end{table} 

The order in which the $2$-element sets are picked here is called co-lexicographic order. For any two subsets $A$ and $B$ of an ordered set, we say $A > B$ in  co-lexicographic order if $\max\big(A\Delta B) \in A$, where $ A \Delta B = (A \setminus B) \cup (B \setminus A)$. For instance we have, $\{1,2\} < \{1,3\}$ in  co-lexicographic order since $3 \in  \{1,3\}$. Hence, $\{1,2\}$, $\{1,3\}$, $\{2,3\}$, $\{1,4\}$, $\{2,4\}$, $\{3,4\}$, $\{1,5\}$, $\{2,5\}$, $\{3,5\}$, $\{4,5\}$  are in co-lexicographic order. %This means that the shortening procedure given in table:~\ref{table:PRM_2_4} picks the message symbols corresponding to first $\gamma$ $2-$element subsets of $[4]$ in co-lexicographic order. 
Although it is not explicitly stated in \cite{VajRamKum}, the general shortening procedure used to prove Lemma~\ref{thm:shorten3} picks first $\gamma$ $r-$element subsets of $[m]$ in co-lexicographic order.\\
The terminology anti-lexicographic order is used for the reverse co-lexicographic order. For example, $\{4,5\}$, $\{3,5\}$, $\{2,5\}$, $\{1,5\}$, $\{3,4\}$, $\{2,4\}$, $\{1,4\}$, $\{2,3\}$, $\{1,3\}$, $\{1,2\}$ are in anti-lexicographic order. Hence, the remaining message symbols after shortening will correspond to the first  $k$ $r-$element subsets of $[m]$ in anti-lexicographic order.

\begin{thm}
	For the binary $\text{PRM}(r, m-1)$ code, the $k$-th generalized Hamming weight
	\bea
	\label{eq:ghw}
	d_k(r,m) \le   \sum\limits_{i = r}^m {m \choose i} - \Gamma(r,m,\gamma), 
	\eea where $\Gamma(r,m,\gamma)$ is the block length reduction given by Lemma~\ref{thm:shorten3} for $\gamma= {m \choose r}-k$.
\end{thm}
\bprf
The shortened version of $\text{PRM}(r,m-1)$ code obtained by setting first $\gamma= {m \choose r}-k$  message symbols in co-lexicographical order to  zero is a $k$-dimensional sub code of the $\text{PRM}(r,m-1)$ code. Therefore, the block length of this shortened code  gives an upper bound on the $k$-th GHW of the $\text{PRM}(r,m-1)$ code. 
\eprf

\ \\
%For example, consider the case of $d_1$ which is also the minimum hamming weight, the $(r,m)$ canonical form is given by $t=1, r_1 = 0, m_1 = m-r$. Substituting this in Corollary~\ref{cor:ghw} we get $d_1 = 2^{m-r}$.
\section{Lower Bound On GHW Of Binary PRM Codes\label{sec:lower}}
\begin{figure}[h!]
	\begin{center}
		\includegraphics[width=4.4in]{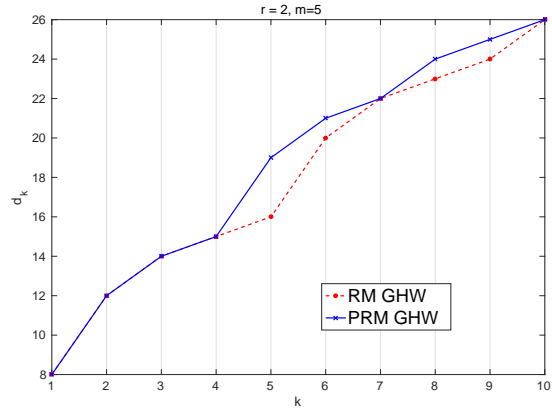}
	\end{center}
	\caption{Gap between GHW of PRM$(2,4)$ and RM$(2,5)$ codes, here $r=2$, $m=5$.}\label{fig:ghw_gap}
\end{figure}

In Theorem \ref{thm:lb_ghw} we present a lower bound on GHW for binary PRM codes. The proof shown here adapts techniques from the proof for Reed-Muller codes in \cite{Wei}.
The GHW for RM codes determined in \cite{Wei} give a lower bound on GHW for PRM codes since PRM codes are subcodes of RM codes with same parameters. However, we will prove that there is gap between GHW of RM and PRM codes (see Figure \ref{fig:ghw_gap}) by proving a tighter lower bound for PRM codes.% it can bshows the gap between GHW of RM and PRM codes for $r=2$, $m=5$. Therefore

Every codeword in PRM$(r,m-1)$ code corresponds to evaluations of a binary homogeneous polynomial of degree $r$ in $m$ variables. Hence, we use the notation $f \in $ PRM$(r,m-1)$ to represent the codeword given by evaluations of homogeneous polynomial $f \in \mathbb{F}_2[x_1, \cdots, x_m]$. It can be seen that any $f \in $ PRM$(r,m-1)$ can be represented as 
$f = f_1 + x_m f_2$, where $f_1 \in $ PRM$(r,m-2)$ and $f_2 \in $ PRM$(r-1,m-2)$. 
\bthm \label{thm:lb_ghw}
For any $0 \le k < {m \choose r}$,
\bean
\scalebox{0.95}{$
d_k(r,m) \ge \min \limits_{\substack{s+t = k\\ s \le {m-1 \choose r-1}, \ t \le {m-1 \choose r}}} \{ d_s(r-1,m-1) + d_t(r,m-1) \}$}
\eean
\ethm
\bprf
Let $\mathcal{C}$ be a subcode of PRM$(r,m-1)$ with support size $d_k(r,m)$ and dimension $k$. 

Let $L = \mathbb{F}_2[x_1, \cdots, x_{m-1}]$. We define,
\bean
\mathcal{C}_1 = \{ f \in L ; \ x_m f \in \mathcal{C} \}.
\eean
Let $\mathcal{C}_2$ be such that $\mathcal{C} = x_m \mathcal{C}_1 \oplus \mathcal{C}_2$, where $\oplus$ denotes direct sum, $x_mC_1 = \{x_m f | f \in C_1\}$. We now define define
\bean
\mathcal{C}_3 = \{g \in L; \exists f \in L, \ x_m f + g \in \mathcal{C}_2\}. 
\eean
Let the dimension of $\mathcal{C}_1$, $\mathcal{C}_2$ be $s^*$, $t^*$ respectively. It can be observed that $\mathcal{C}_1$ is a subcode of PRM$(r-1,m-2)$ and $\mathcal{C}_3$ a subcode of PRM$(r,m-2)$. Therefore, any element in $\mathcal{C}_2$ can be written as $x_m f + g$, where  $f,g \in L$ are homogeneous polynomials and $deg(f) = r-1$, $deg(g) = r$. 

We will now show that $\mathcal{C}_3$ and $\mathcal{C}_2$ have same dimension. If $g \in \mathcal{C}_3$, then there exists $f_1 \in L$ such that $x_m f_1 + g \in \mathcal{C}_2$. If there is $f_2 \ne f_1$ such that $x_m f_2 + g \in \mathcal{C}_2$, it would imply that $x_m (f_1 + f_2) \in \mathcal{C}_2$. But $x_m (f_1 + f_2) \in x_m \mathcal{C}_1$, resulting in a contradiction. Therefore for every element in $\mathcal{C}_3$, there is a corresponding unique element in $\mathcal{C}_2$. %It can also be checked that $\mathcal{C}_1 \cap \mathcal{C}_3 = \{0\}$.

The support size of subcode $\mathcal{C}$ is given by
\bean
d_k(r,m)=|S(\mathcal{C})| = |S_0(\mathcal{C})| + |S_1(\mathcal{C})|,
\eean
where $S_i(\mathcal{C})$ corresponds to support when $x_m = i$ for $i = 0,1$. Since $\mathcal{C}_1$ is a $s^*$ dimensional subcode of PRM$(r-1,m-2)$, we have $|S(\mathcal{C}_1)|\ge d_{s^*}(r-1,m-1)$ and similarly $|S(C_3)| \ge d_{t^*}(r,m-1)$.\\
It is clear to see that $|S_1(\mathcal{C})| \ge |S(\mathcal{C}_1)| \ge d_{s^*}(r-1,m-1)$ and $|S_0(\mathcal{C})| \ge |S(\mathcal{C}_3)| \ge d_{t^*}(r,m-1)$.
Thus we have, 
\bean
d_k(r,m) &\ge& d_{s^*}(r-1,m-1) + d_{t^*}(r,m-1)\\
&\ge& \scalebox{0.9}{$\min\limits_{\substack{s+t = k\\ s \le {m-1 \choose r-1}, \ t \le {m-1 \choose r}}} \{ d_s(r-1,m-1) + d_t(r,m-1) \}$}
\eean
\eprf

%\begin{note}
%	In the minimization over any $s+t=k$, the parameters $d_s(r-1,m-1)$ and $d_t(r,m-1)$ are defined only for $s < {m-1 \choose r-1}$ and ${m-1 \choose r}$ respectively.
%\end{note}
\section{GHW Of Binary PRM Codes\label{sec:ghw}}

In this Section, we will first state a well-known theorem from extremal set theory and then use it to prove the GHW results. %We use ideas from \cite{Wei} in proofs in this section as well.

Let $U_r$ denote the family of all $r$-element subsets of $[m]$. For a collection $K \subseteq U_r$, (upward) shadow is given by
\bean
\Delta(K) = \{ X \subseteq [m] \ | \ Y \subseteq X \ \text{ for some } Y \in K \}
\eean

\bthm[Kruskal \cite{Kruskal}, Katona \cite{Katona} ]The collection $K$ consisting of first $k$ $r$-subsets of $[m]$ picked in anti-lexicographic order achieves $\min \{|\Delta(K)\cap U_{r+1}|: K \subseteq U_r$ and $|K| = k \}$.
\ethm
\bcor\label{cor:minwt_antilex}
The collection $K$ consisting of first $k$ $r$-subsets of $[m]$ picked in anti-lexicographic order achieves $\min \{ |\Delta(K)|: K \subseteq U_r$ and $|K| = k \}$.
\ecor
 
Let $\sigma_{(r,m,k)}$ denote the support size of subcode formed by monomials corresponding to first $k$ $r$-element subsets of $[m]$ in the anti-lexicographic order. Note that this subcode is same as the shortened PRM code formed by setting ${m \choose r} - k$ message symbols picked in co-lexicographic order to zero. Since this subcode is $k$ dimensional subcode of PRM$(r,m-1)$ we have $\sigma_{(r,m,k)} \ge d_k(r,m)$. \\

%{\color{red}\begin{note} Avoid this ???
%	Note that, for any $0 \le k < {m \choose r}$,
%	\bean
%	\sigma_{(r,m,k)} = \sum\limits_{i=r}^m {m \choose i} - \Gamma\Big(r, m, {m \choose r} - k\Big).
%	\eean
%\end{note}}

\blem
\label{lem:wt_antilex}
\bean
\sigma_{(r,m,k)} \le \min \limits_{\substack{s+t = k\\ s \le {m-1 \choose r-1}, \ t \le {m-1 \choose r}}} \{ \sigma_{(r-1,m-1,s)} + \sigma_{(r, m-1,t)} \}
\eean
\elem
\bprf
Suppose $s^*, t^*$ be the values that achieve minimum on the RHS.
Then RHS $= \sigma_{(r-1,m-1,s^*)} + \sigma_{(r,m-1,t^*)}$. The RHS corresponds to the support size of subcode formed by $k$ monomials each of degree $r$, with $s^*$ of them containing $m$ and $t^*$ of them without $m$. Now, from Cor \ref{cor:minwt_antilex} we know that picking the first $k$ $r$-element sets in anti-lexicographic order results in minimum support size $\sigma_{(r,m,k)}$. Hence, RHS $\ge \sigma_{(r,m,k)}$.
\eprf

\bthm \label{thm:ghw_prm}
For any $0 \le k < {m \choose r}$,
\bean 
\sigma_{(r,m,k)} = d_k(r,m)
\eean
The support of subcode formed by first $k$ monomials in anti-lexicographic order is the $k$th GHW of PRM$(r,m-1)$.
\ethm
\bprf
For the case of $m=1$, the statement trivially follows. Now, we use  induction over $m$ to prove the theorem and hence assume it is true for $m-1$.

Let $\mathcal{C}$ be the subcode with rank $k$ and support size $d_k(r,m)$. Then by Theorem \ref{thm:lb_ghw} we have,
\bea
|S(\mathcal{C})| &\ge& \scalebox{0.92}{$\min \limits_{\substack{s+t = k\\ s \le {m-1 \choose r-1}, \ t \le {m-1 \choose r}}} \{ d_s(r-1,m-1) + d_t(r,m-1)\}$} \nonumber \\
&=& \min \limits_{\substack{s+t = k\\ s \le {m-1 \choose r-1}, \ t \le {m-1 \choose r}}} \{ \sigma_{(r-1,m-1,s)} + \sigma_{(r,m-1,t)}\} \label{eq:ind}\\
&\ge& \sigma_{(r,m,k)} \label{eq:wt_antilex}\\
&\ge& d_k(r,m) \nonumber \\
&=& |S(\mathcal{C})| \nonumber
\eea
Here, \eqref{eq:ind} follows from induction assumption and \eqref{eq:wt_antilex} follows from Lemma \ref{lem:wt_antilex}.
Therefore, $\sigma_{(r,m,k)} = d_k(r,m)$.
\eprf
\bcor
\label{cor:ghw}
For any $0 \le k < {m \choose r}$,
\bean
d_k(r,m) = \sum\limits_{i=r}^m {m \choose i} - \Gamma\Big(r, m, {m \choose r} - k\Big)
\eean
where $\Gamma$ is obtained from Theorem \ref{thm:shorten3}
\ecor
\bprf
From Theorem \ref{thm:ghw_prm}, support of $k$ monomials picked in anti-lexicographic order gives the $k$-th generalized Hamming weight. This is same as avoiding (shortening) the first ${m \choose r} - k$, monomials picked in co-lexicographic order. Therefore, the support size obtained from  Theorem \ref{thm:shorten3} with $\gamma={m \choose r} - k$ is equal to $\sigma_{(r,m,k)}$ and hence the result.
\eprf

The above corollary proves the optimality of shortening procedure for PRM codes given in \cite{VajRamKum} and provides the complete GHW hierarchy. GHW hierarchy of PRM codes for some parameters are listed in Table \ref{table:PRM_ghw}. The next corollary gives simplified expression for GHW in some special cases. 
\bcor
\bean
d_k(r,m) = (2^k - 1)2^{m-r-k+1} \ \ \text{ for } k \le m-r+1.
\eean
\ecor
\bprf
Pick first $k$, $r$-element subsets of $[m]$ in anti-lexicographic order. These sets are of the form:
\bean
S_i = \{m, m-1, \cdots, m-r+2, m-r-(i-2) \},
\eean
for all $i \in [k]$. 

Now consider the monomials corresponding to $S_i$, for all $i \in [k]$. %and the space $K$ spanned by the evaluation of these monomials. 
The vector $\underline{x} \in \mathbb{F}_2^m \setminus \{\underline{0}\}$ for which at-least one of these monomials evaluate to one has, $x_i = 1$ for all $i \in [m-r+2,m]$ and  $x_i=1$, for at-least one $i \in [m-r-k+2,m-r+1] $. The remaining $x_i, i \in [1,m-r-k+1]$ can take any value.
The number of such vectors is $(2^k - 1)2^{m-r-k+1}$.
\eprf

\begin{note}
	\bean
	d_1(r,m) &=& 2^{m-r},	\\
	d_2(r,m) &=& 3\cdot2^{m-r-1}; \ \	m \ge r+1,\\
	d_3(r,m) &=& 7\cdot2^{m-r-2}; \ \	m \ge r+2.
	\eean
\end{note}
The GHW for PRM code obtained in Cor \ref{cor:ghw} is by considering the sets to be removed. In the GHW derivation for Reed-Muller codes in \cite{Wei}, the counting is done taking into account the sets that remain. The following lemmas gives an expression for GHW of PRM codes using a similar approach.
\blem
Any $0 \le k < {m \choose r}$ can be uniquely represented by $(r,m)$ canonical form given by
\bean
k = \sum\limits_{i=1}^t {m_i \choose r_i}
\eean
where, $r > r_1 \ge r_2 \cdots \ge r_t \ge 0$, $m_i \ge 0$ and $m_i-r_i = m-r-i+1$.
\elem
\bprf
We induct over variable $m$.

For $m=1$, the result is trivial. Assume that the statement true for $m-1$.

If $k \ge {m-1 \choose r-1}$, define $k' = k - {m-1 \choose r-1} < {m-1 \choose r}$. Then, $k'$ has a $(r, m-1)$ canonical representation given by:
\bean
k' = \sum\limits_{i = 1}^{t'} {m_i' \choose r_i'}
\eean
Setting $m_1 = m-1$, $r_1 = r-1$ and $m_{i+1} = m_i'$ and $r_{i+1} = r_i'$ for all $i \in [t']$ satisfies the lemma statement with $t=t'+1$.\\
For $k < {m-1 \choose r-1}$, the $(r-1, m-1)$ canonical form will itself be the $(r,m)$ canonical form.
\eprf
\ \\
\bcor\label{cor:ghw}
For any $0 \le k < {m \choose r}$, the k-th GHW of binary projective Reed Muller code PRM$(r,m-1)$ is given by:
\bean 
d_k(r,m) = \sum\limits_{i=1}^t \sum\limits_{j = r_i}^{m_i} {m_i \choose j}.
\eean
\ecor

\bprf
Here, we induct on variable $m$. Assume that the result holds for the case of $m-1$.

Let $K$ be the set of first $k$, r-element subsets of $[m]$ in anti-lexicographic order.

 Consider the case of $k \ge {m-1 \choose r-1}$, here $K$ exhausts all the $r$-element subsets that include $m$. Suppose $S_1(K)$ represents support generated by sets that include $m$ and $d_{k'}(r,m-1)$ the support generated by remaining sets in $K$, where $k' = k - {m-1 \choose r-1}$. Then, 
 \bean
 |S(K)| = |S_1(K)| + d_{k'}(r,m-1). 
 \eean 
 It can be observed that $|S_1(K)|$ is same as the block length of  PRM$(r-1,m-2)$ code.
Therefore by induction assumption, 
\bean
|S(K)| = \sum\limits_{j = r-1}^{m-1} {m-1 \choose j} + \sum\limits_{i=1}^{t'} \sum\limits_{j=r_i'}^{m_i'} {m_i' \choose j}
\eean 
where $(m_i', r_i'), \ \forall i \in [t']$ is the $(r,m-1)$ canonical representation of $k'$.

Now by picking $m_1 = m-1$, $r_1 = r-1$, $m_{i+1} = m_i'$, $r_{i+1} = r_i'$ for all $i \in [t']$ and $t=t'+1$, we get:
\bean
|S(K)| &=& \sum\limits_{i=1}^{t} \sum\limits_{j=r_i}^{m_i} {m_i \choose j}\\
&=& d_k(r,m)
\eean
The second equality follows from Theorem \ref{thm:ghw_prm}.

For the case of $k < {m-1 \choose k-1}$, all the $r$-element sets in $K$ include $x_m$ and the support generated by these sets can therefore be determined by $d_k(r-1,m-1)$. The $(r-1,m-1)$ canonical representation for $k$ is also $(r,m)$ canonical representation for $k$.
\bea
d_k(r,m) &=& d_k(r-1,m-1) \nonumber \\
&=& \sum\limits_{i=1}^t \sum\limits_{j=r_i}^{m_i} {m_i \choose j} \label{eq:dkind}.
\eea
Equation \eqref{eq:dkind} follows by induction assumption.
\eprf

\begin{table}[h!]
	\begin{center}	
		\bean
		\begin{array}{|c|c|c|}
			\hline r & m  & \text{ GHW Hierarchy } \Big(d_1, \cdots, d_{m \choose r}\Big)\\ \hline
			1 & 2 & 2,3 \\ \hline
			1 & 3 &  4,6,7 \\ \hline
			2 & 3 &  2,3,4  \\ \hline
			1 & 4 &  8,12,14,15 \\ \hline
			2 & 4 & 4,6,7,9,10,11 \\ \hline
			3 & 4 & 2,3,4,5 \\ \hline
			1 & 5 & 16,24,28,30,31 \\ \hline
			2 & 5 & 8,12,14,15,19,21,22,24,25,26  \\ \hline
			3 & 5 & 4,6,7,9,10,11,13,14,15,16 \\ \hline
			4 & 5 & 2,3,4,5,6 \\ \hline
		\end{array}
		\eean
		\caption{ GHW hierarchy of binary PRM$(r,m-1)$ code for some parameters.}
		\label{table:PRM_ghw}
	\end{center}
\end{table} 
%\section{Conclusions}
%This paper focuses on determining the GHW fot binary Projective Reed-Muller codes. The GHW for Projective Reed-Muller codes has been extensively studied in the past, but most of those works focused on the $r<q$ regime. The specific case of $q=2$ (binary) has never been considered. Here, we show that Victor Wei's derivation of GHW for binary Reed-Muller codes in \cite{Wei} can be extended to binary PRM codes with appropriate modifications. 
\bibliographystyle{IEEEtran}
\bibliography{ncc}
\end{document}